\newcommand{\lmax}{\lambda_{\text{max}} }

\documentclass[twocolumn,pre,amsmath,amssymb,superscriptaddress]
              {revtex4}
\usepackage{bm,color,nicefrac}
\usepackage[tight]{subfigure}
\usepackage[pdftex]{graphics}
\usepackage{epsfig}
\usepackage{amssymb}
\usepackage{amsmath}
\usepackage{bm}
\usepackage{xcolor}
\usepackage{float}
\definecolor{TangoChameleon3}{HTML}{4E9A06}
\definecolor{TangoSkyBlue2}{HTML}{3465A4}
\definecolor{TangoScarletRed2}{HTML}{CC0000}
\definecolor{TangoAluminium2}{HTML}{D3D7CF}

\newcommand{\oskar}[1]{\textcolor{TangoSkyBlue2}{}}

\begin{document}

\title {Thermalized connectivity networks of jammed packings}

\author{Clemens Buss} %\email{clemens.buss@ds.mpg.de}
\affiliation{Biophysics and Evolutionary Dynamics Group, Max Planck Institute for Dynamics and Self-Organization, Am Fa{\ss}berg, 37077 G{\"o}ttingen, Germany}
\affiliation{Institute for Theoretical Physics, Georg-August University of G{\"o}ttingen, Friedrich-Hund Platz 1, 37077 G{\"o}ttingen, Germany}
\author{Claus Heussinger} 
\affiliation{Institute for Theoretical Physics, Georg-August University of G{\"o}ttingen, Friedrich-Hund Platz 1, 37077 G{\"o}ttingen, Germany}
\email{heussinger@theorie.physik.uni-goettingen.de}
\author{Oskar Hallatschek} \email{ohallats@berkeley.edu}
\affiliation{Biophysics and Evolutionary Dynamics Group, Departments of Physics and Integrative Biology, University of California, Berkeley, CA 94720} \date{\today}
\begin{abstract}
  Jammed packings of repulsive elastic spheres have emerged as a
  rich model system within which elastic properties of disordered
  glassy materials may be elucidated. Most of the work on these
  packings have focused on the case of vanishing temperature. Here, we
  explore the elastic properties of the associated connectivity
  network for finite temperatures, ignoring the breaking of bonds and
  the formation of new ones. Using extensive Monte Carlo simulations,
  we find that, as the temperature is increased, the resulting spring
  network shrinks and exhibits a rapidly softening bulk modulus via a
  cusp. Moreover, the shear modulus stiffens in a fixed volume
  ensemble but not in a fixed pressure ensemble. These
  counter-intuitive behaviors may be understood from the characteristic
  spectrum of soft modes near isostaticity, which resembles the
  spectrum of a rod near its buckling instability. Our results suggest a
  generic mechanism for negative thermal expansion coefficients in
 marginal solids. We discuss the consequences of bond breaking and an apparent analogy between thermalization and shear. 
\end{abstract}
\maketitle

% \oskar{To improve intro, we may revisit dennisson/mackintosh, Tighe, Mathieu/Maha, Lubensky et al.}

Packings of repulsive elastic particles have emerged as a rich model
granular system with potential relevance to amorphous solids~\cite{liu1998nonlinear,liu2010dynamical}. A
number of remarkable features emerge, in particular, when the packing
is close to the onset of rigidity. One of the key characteristics of jammed packings is  an excess of low-frequency vibrations~\cite{silbert2005vibrations}, the so-called boson peak. Numerous consequences can be derived from this peculiar vibrational spectrum,
with regards to, for instance, elastic or transport properties~\cite{o2003jamming,wyart2005effects,xu2009energy}. Some of these features are shared with lattices close to
isostaticity~\cite{wyart2008elasticity,ellenbroek2009non,tighe2011relaxations,dennison2013fluctuation,mao2015mechanical,wigbers2015stability}, which may be exploited to develop meta-materials with
novel mechanical
properties~\cite{kane2014topological,paulose2015topological}.  

While most studies have focused on the zero-temperature consequences
of the vibrational spectrum, we here study the impact of thermal
fluctuations. Specifically, we consider the harmonic connectivity network obtained from a jammed packing of repulsive, friction-less spheres close to
isostaticity, and study its mechanical properties as we heat up the system to  a low but finite temperature. 

Elastic properties of ordered and disordered networks of springs at finite temperatures have been studied previously~\cite{boal1993negative,discher1997phase,tessier2003networks}. Most recently, motivated by the attractive properties of highly responsive marginal solids for material science and biophysics, spring networks have been studied near the isostatic threshold~\cite{dennison2013fluctuation,mao2015mechanical,wigbers2015stability,feng2016nonlinear}. These studies revealed, amongst others, interesting anomalies in the entropic elasticity. While these studies have focused on networks that have soft bulk and shear moduli, as in rigidity percolation~\cite{bolton1990rigidity}, it is a characteristic of jammed networks, studied in this work, to have a finite bulk modulus at isostaticity~\cite{ellenbroek2009non}. As we will see, this has major consequences for the impact of thermal fluctuations on the material properties of the network. 

{\bf Simulation approach.} To prepare the initial conditions for our simulations, we generate jammed packings of repulsive elastic spheres. Each packing is created through an energy minimization protocol with thermal equilibration, as described in
Ref.~\cite{o2003jamming}. The protocol results in a series of contact
networks at different coordination numbers above a critical value
$z_c=4$.  \oskar{Clemens, you also need to mention and cite the protocol of going through cycles of inflation deflation to remove hysteresis and make starting point well-defined.}

The minimal contact number $z_c$ per particle required for rigidity
follows from Maxwell's counting argument~\cite{maxwell1864calculation}. Since there are $d$ degrees
of freedom for a point particle, one needs at least $z_c=2d$ contacts
to constrain the positions of all particles. Networks with $z_c$
contacts are called isostatic.

Our hyperstatic contact networks ($z>z_c$) are then modeled as a
network of Hookian springs. The effect of thermal fluctuations is
studied using Metropolis Monte Carlo simulations. Due to the small temperatures
considered, crossing of bonds is very unlikely, although it is not
penalized in our simulations. Also, contacts do neither break
nor form in the course of the simulation. Consequences of bond breaking are discussed below.

\section{ Simulation Results}
At fixed tension (negative pressure), we
observe that our spring networks contract linearly with increasing
temperature, as was found previously in high-coordination number networks~\cite{boal1993negative}. Importantly, we find that the contraction diverges in a characteristic way as the connectivity approaches isostaticity. 

Specifically, we find that the negative thermal expansion coefficient, NTE, exhibits a singularity as the average contact number per particle approaches isostaticity.  For vanishing tension $\tau$, we find that the NTE scales inversely with the distance $\delta z=z-z_c$ from the critical contact number. For large tensions, on the other hand, the
NTE$\sim \tau^{-1/2}$ is independent of contact number. Fig. 1 shows
that the data for different $T$, $\delta z$ and $\tau$ collapse onto a
master curve when we plot NTE $\delta z$ vs. $\tau \delta z^{-2}$. The deviations
suggest that the collapse for larger tensions only works for small
enough $\delta z$.

Like jammed packings, our spring networks are affinely deformed when compressed at zero temperature. In other words, compressive
deformations merely change contact lengths but not  contact
angles. Consequently, the bulk modulus $B_0$ of zero-temperature
jammed networks is set simply by the spring constant and the co-ordination number.

By contrast, as we turn on temperature, we see that the material
becomes much softer to compression, which is manifest by a cusp in the
bulk modulus. Specifically, the difference $B^{-1}-B_0^{-1}$ scales as
$T \delta z^3$ for vanishing tension, Fig.~\ref{Fig:BulkScaling}. As we increase the tension on
the network, the behavior crosses over to being proportional to
$T/\tau^{3/2}$.

The shear modulus is inconspicuous for small enough tension, as it
follows the zero temperature result $G\sim \delta z$, which is small
near the critical point due to soft modes. For fixed area, however,
the shear modulus crosses over to $G\sim T^{1/3}$ once the product
$T \delta z^3$ becomes of order unity, as shown in Fig.~\ref{Fig:shearmodulus}. This behavior is markedly different from the scaling $G\sim T^{1/2}$ observed in disordered spring networks near isostaticity~\cite{dennison2013fluctuation}.

\begin{figure}
  \centering
  \includegraphics[width=\columnwidth]{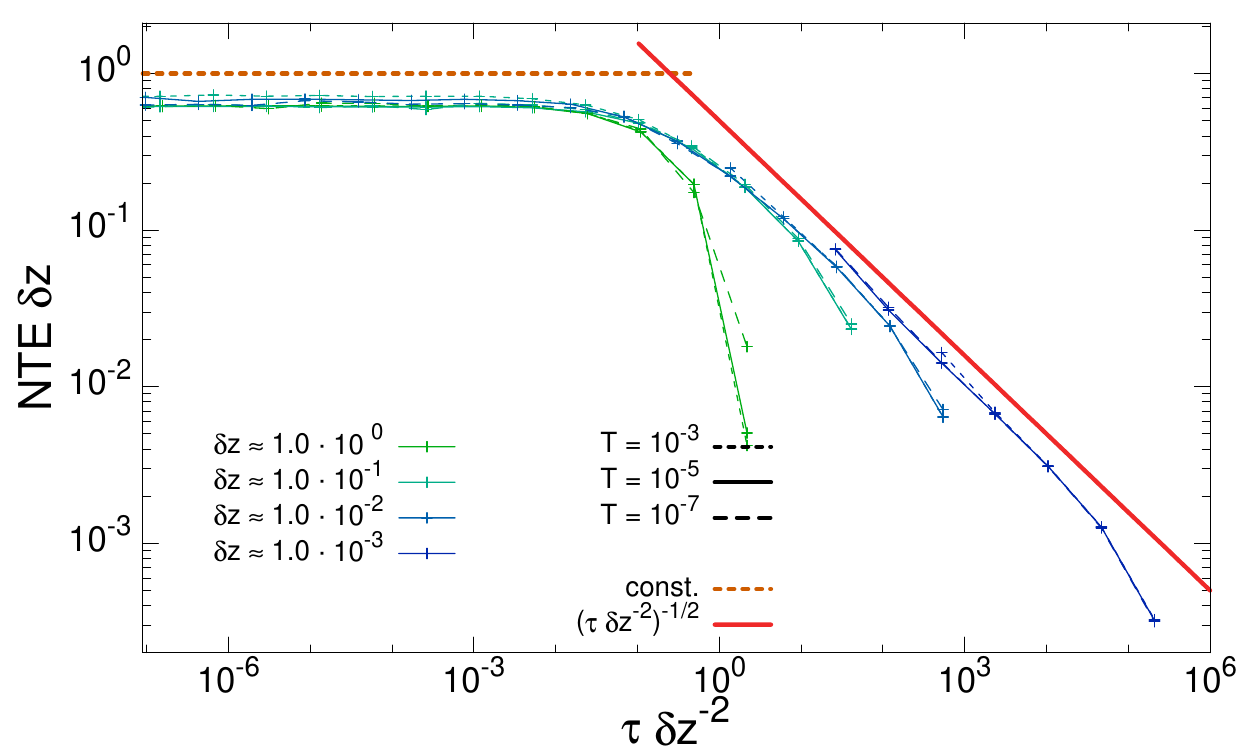}
  \caption{Scaled Negative thermal expansion coefficient (NTE) as a
    function of rescaled tension $\tau \delta z^{-2}$ for various
    combinations of temperature and coordination number difference
    $\delta z$ from criticality. The dashed and solid red lines
    represent our scaling predicitions for small and large
    tension. The network size is $N=2000$.}
  \label{Fig:AreaScaling}
\end{figure}

\begin{figure}
  \centering
  \includegraphics[width=\columnwidth]{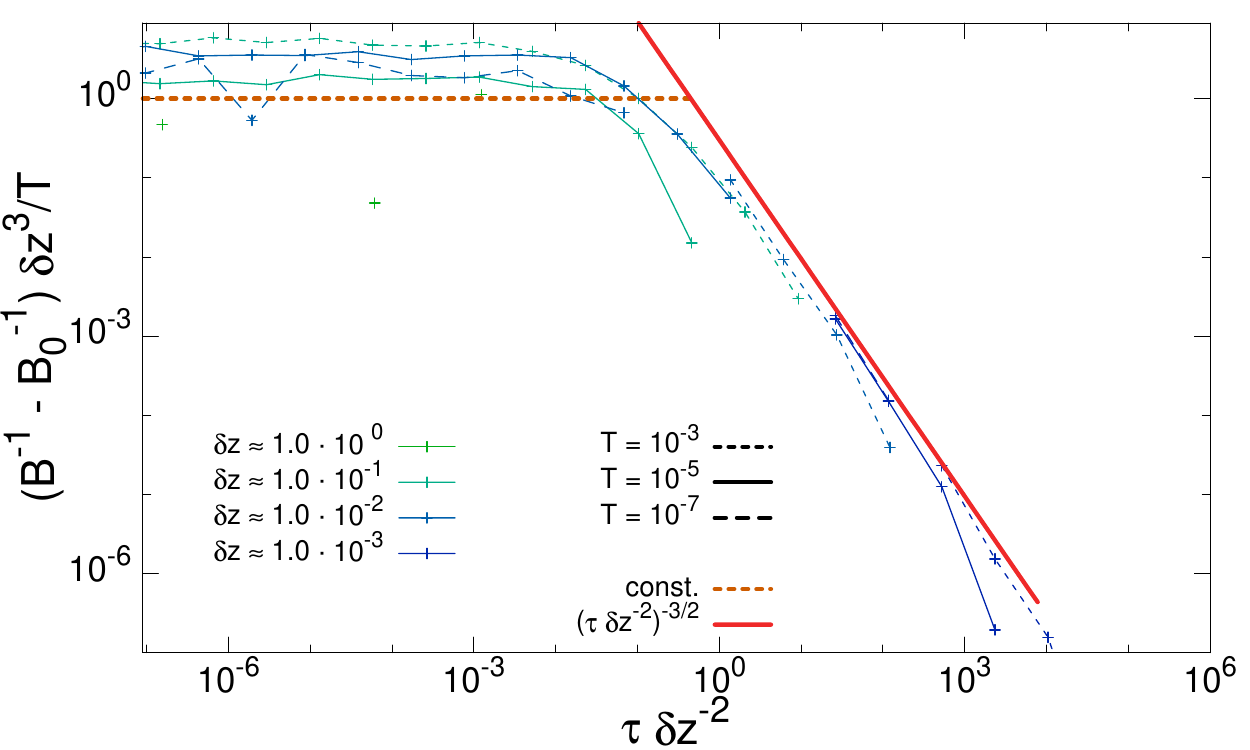}
  \caption{The scaled difference of inverse bulk modulus at finite
    temperature and tension, $B^{-1}$, and at zero temperature and
    tension modulus, $B_0^{-1}$, on a double logarithmic scale as a
    function of scaled tension, $\tau \delta z^{-2}$. Different data
    sets correspond to different combinations of temperature and
    $\delta z$, as indicated in the legend. The dotted and solid
    orange line indicates our scaling predictions for small and large
    tension, respectively. The network size is $N=2000$. }
  \label{Fig:BulkScaling}
\end{figure}

\begin{figure}
  \centering
  \includegraphics[width=\columnwidth]{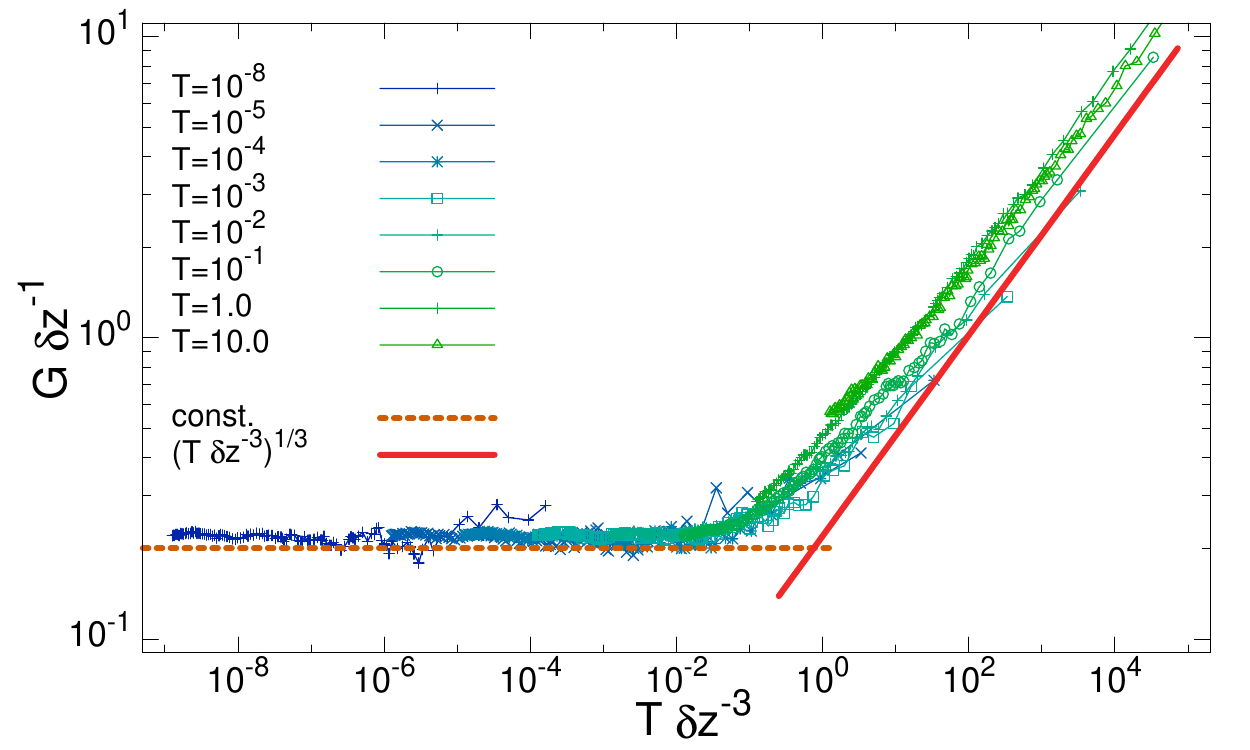}
  \caption{Scaled shear modulus on a double logarithmic scale as a function of scaled temperature $T \delta z^{-3}$.
    Different data sets correspond to different combinations of temperature and $\delta z$, as indicated in the legend.
    The coordination $\delta z$ is ranged from $10^{-2}$ to $1.99$ in $100$ log-spaced steps.
    The dashed and solid lines indicate our scaling predictions for small and large temperature, respectively.
    The network size is $N=1600$. }
  \label{Fig:shearmodulus}
\end{figure}

% --------------

% \begin{equation}
%   \label{eq:scaling-form-NTE}
%   NTE\sim \delta z^{-1} \hat NTE\left(\tau \delta z^{-2}\right)
% \end{equation}

In summary, we find that jammed networks contract upon heating and,
respectively, become more tensed at fixed volume. As a consequence,
the bulk modulus softens for fixed pressure and the shear modulus
hardens for fixed volume.

\section{Intuitive picture based on a square lattice.} We now show that these properties are a consequence of the peculiar vibrational spectrum of jammed networks.

\begin{figure}[!tb]
\begin{center}
\includegraphics[width=0.5\textwidth]{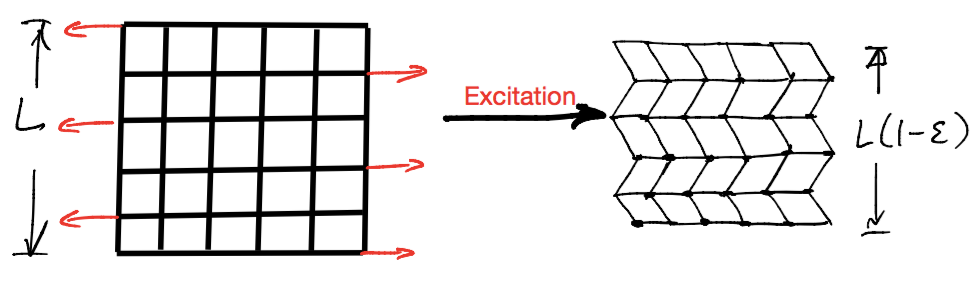}
\caption{\label{fig:square}
\textbf{Effect of floppy mode fluctuations on a square lattice.}
(Left) A square lattice of linear dimension $L$ at vanishing temperature. The red arrows indicate one of the $L^{d-1}$ soft modes of the lattice. (Right) Excitation of this particular soft mode, for instance by thermal fluctuations or shear stress, leads to the contraction of the vertical end-to-end distance.  In this state, an external tension will pull out the zig-zag undulations rather than stretch individual bonds. These features are generic to jammed networks, which likewise contract upon excitation as we argue based on the known mode spectrum. }
\end{center}
\end{figure}

The basic physics can be understood by considering the square lattice
in Fig.~\ref{fig:square}. Despite being ordered, this isostatic lattice shares the
vibrational properties of jammed networks at criticality. Soft modes
are easily identified as zig-zag modes that can be excited at zero
tension without energy cost. There are $O(N^{1/2})$ such zig-zag
modes, one for each boundary node. The deformations induced by one of
these soft modes are indicated by arrows in the figure.

Given this cartoon picture of an isostatic network, an important
observation can be made that may, at first, seem special to the square
lattice: Exciting the said zig-zag modes turns straight
system-spanning lines into broken lines of shorter end-to-end
distance. As a consequence, thermally exciting many such modes leads
to a collapse of total network area (volume in 3D).

A finite negative thermal expansion coefficient is obtained when one
rigidifies the lattice, which can be done in various ways. Here, we
focus on adding $\delta z$ extra contacts per node and on applying a
tension $\tau$. The erstwhile zero-frequency modes now stiffen in a
characteristic way. As
a consequence, the negative thermal expansion coefficient acquires a
finite value that diverges as a powerlaw as $\delta z$ and the tension
vanish. The feature of straight lines reducing their end-to-end
distance by thermal undulations is reminiscent of the physics of 
polymers~\cite{doi1988theory,rubinstein2003polymers}. Indeed, we will see that
the mode spectrum of near isostatic networks and semiflexible polymers
has striking similarities.  

Pulling the boundaries of the non-excited square lattice, i.e. one
applies a tension, obviously stretches the springs in a simple affine
way. This explains why the zero-temperature bulk modulus is given by
the spring constant. At finite temperature, however, soft modes are
excited and the lattice is characterized by jagged lines. An external
stretching force now can pull out undulations of these jagged lines rather than affinely
stretching bonds. The bulk modulus, thus, softens due to the presence
of undulations that can be pulled out. This phenomenon is equivalent
to the longitudinal response of semiflexible polymers being finite
only at finite temperatures, where undulations exist that can be
pulled out by external forces~\cite{doi1988theory,rubinstein2003polymers}.

At fixed area, exciting soft modes requires stretching of bonds. As a
consequence, the tension in the networks increases. This increase in
tension stabilizes the network by generating a finite shear modulus at isostaticity.

Next we turn these intuitive arguments into scaling arguments to show
that, actually, the above behavior is not special for square lattices
but characteristic for the harmonic response of jammed
packings. Afterwards we will discuss the relation of our results to
random networks generated in other ways, for instance by randomly
cutting excess bonds.

\section{Scaling Arguments}
\label{sec:scaling-arguments}
Our scaling analysis is based on an estimate of how the positional fluctuations of the nodes in the network depend on temperature and tension. These fluctuations can then be used to estimate the negative thermal expansion coefficient and the elastic moduli.

We assume that the spring constant $k$ and rest length $a$ of the
springs (at vanishing temperature and pressure) is identical for all
springs. Thus, it is convenient to measure lengths in units of $a$,
energies in units of $ka^2$, the temperature $T$ in units of $k a^2/k_\text{B}$ and the tension in units of $k$. Equivalently, we set $a=1$ and $k_\text{B}=1$.

Since our goal is to explain the observed scaling laws, we will not keep track of
numerical pre-factors of order unity.

\subsection{Mean Square Displacement}
\label{sec:MSD}
We consider a large hyperstatic $d$-dimensional spring network with co-ordination number $z_c+\delta z$ above the minimal isostatic value $z_c=2d$ needed for rigidity~\cite{maxwell1864calculation}. Within such a network, there are of order $\lambda^{d-1}$ springs at the
boundary of a region of linear dimension $\lambda$. Cutting those
boundary springs removes $\lambda^{d-1}$ constraints. If this number of relieved constraints is larger than the excess number $\delta z \lambda^d$ of  constraints
(i.e. springs) we can expect floppy modes in the considered region. Thus, floppy modes extend up to a characteristic length scale $\lmax\sim \delta z^{-1}$~\cite{wyart2005effects}. 

In order to excite a floppy mode of extension $\lambda$ such that transverse
displacements of order $u$ are generated one has to supply energy of
roughly $e_u\sim \lambda^{d} \omega_\lambda^2 u^2$ in terms of the mode frequency
\begin{equation}
  \label{eq:6}
  \omega_\lambda^2=\lambda^{-2}+ c_1\,\tau  
\end{equation}
of a floppy mode of extension $\lambda$~\cite{wyart2005effects} ($c_1$ is a positive constant of order unity). For vanishing tension, $\tau=0$, the above energy
arises because $\lambda^d$ springs are each deformed by a strain of order
$u/\lambda$. The dependence on the tension follows from the fact
that a broken line reduces its end-to-end distance roughly by a
fraction $(u/a)^2/2\sim u^2$ when excited by a transverse displacement
of characteristic wave length $a=1$, as is typical for soft modes
(see, e.g., the soft mode excitation in Fig.~\ref{fig:square}). For a careful derivation of the mode
spectrum, see Ref.~\cite{wyart2005effects}.

\begin{figure}
  \centering
  \includegraphics[width=\columnwidth]{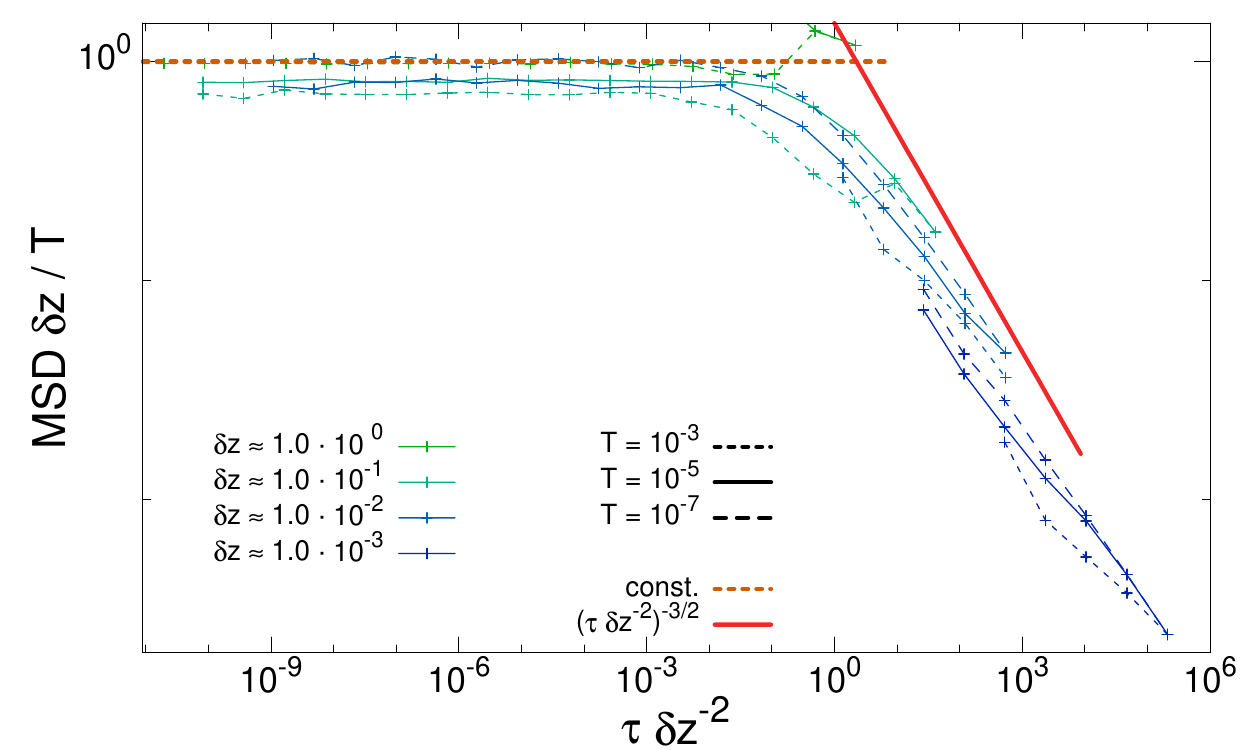}
  \caption{Scaled total mean square displacement as a function of
    $\tau \delta z^{-2}$ for combinations of two temperatures and
    three different $\delta z$. The asymptotics derived from our
    scaling arguments are plotted on top of the data. }
  \label{Fig:MSD}
\end{figure}

At equilibrium, each mode should store an energy of
$k_B T/2$ by the equipartition theorem. Thus, we expect that each floppy mode acquires a mean square amplitude of
$\langle u^2\rangle\sim T \lambda^{-d} \omega_\lambda^{-2}$. The total mean square displacement $\langle\delta R^2\rangle$
of a node in the network can be expected to be dominated by the transverse
(non-affine) displacements caused by the excitation of soft
modes. To estimate the resulting fluctuations, we need to sum up the mean square displacements generated by all orthogonal floppy modes that have overlap with a
given focal node. Since there are $\lambda^{d-1}\sim \int_1^{\lambda}D\left(\tilde\lambda\right)d\tilde\lambda$ floppy modes of extension $\lambda$ overlapping with a given node, we have a density
$D(\lambda)\sim \lambda^{d-2}$ of floppy modes per unit ``wave
length''. The mean square displacement of the focal node is thus given
by the integral
\begin{eqnarray}
  \label{eq:9}
  \langle\delta R^2\rangle&\sim&  \int_1^{\lmax}d\lambda\,
                                 D(\lambda) \frac{T}{ \lambda^{d}
                                 \omega_\lambda^{2}}\\
&\sim&T\int_1^{\lmax}d\lambda\, \left(1+c_1 \tau
       \lambda^{2}\right)^{-1} \\
&=&\arctan\left[\sqrt{c_1 \tau}\lmax\right] (c_1 \tau)^{-1/2} \\
&\sim&\left\{T \delta z^{-1} \qquad \tau\delta z^{-2}\ll1\atop T
       \tau^{-1/2} \qquad \tau\delta z^{-2}\gg1 \right.
\end{eqnarray}
where we used $\lmax\sim\delta z^{-1}$ in going to the last line. Note that these predictions are in agreement with our data collapse in Fig.~\ref{Fig:MSD}.

\subsection{Contraction}
\label{sec:contraction}
Next we study how the (negative)
expansion $\delta A$ of the network area $A$ depends on temperature and
tension. Two contributions have to be considered. The first
contribution is an  affine expansion of network bonds in response to a finite tension. This naive contribution is positive but it turns out to be subdominant compared to the negative non-affine contribution at finite temperature and close to the critical point.

Since in an affine deformation all springs are stretched equally, we can estimate the affine ("stiff") contribution  to the bulk
modulus simply by  $B_a\sim z$, the average number of springs per node. Equivalently the affine relative extension $\epsilon_a$ is given by 
$\epsilon_a=\tau/B_a\sim \tau /z$ in terms of the tension $\tau$.

The non-affine contraction is caused by  the excitation of floppy modes. As we have mentioned above, one excited floppy mode generates a contraction of order
$\langle (u/a)^2/2\rangle\sim \langle u^2\rangle$. Thus, the
 non-affine contraction $-\epsilon_{na}\sim \langle \delta R^2\rangle$
scales the same way as the mean square nodal displacements.

We expect the total extension to be given  by the sum of the
affine and non-affine contributions,
\begin{equation}
  \label{eq:11-0}
  \epsilon=\epsilon_{a}+\epsilon_{na}\sim \frac{\tau}{z}-c_2\,\langle \delta
  R^2 \rangle\;.
\end{equation}
where we introduced another factor $c_2$ of order unity. The negative thermal
expansion  coefficient follows from differentiation with respect to
temperature,
\begin{equation}
  \label{eq:11}
  \text{NTE}=-\partial_T\epsilon\sim\left\{ \delta z^{-1} \qquad
    \tau\delta z^{2}\ll1\atop  
       \tau^{-1/2} \qquad \tau\delta z^{2}\gg1 \right.\;.
\end{equation}
The NTE is shown in Fig.~\ref{Fig:AreaScaling} and exhibits the two
asymptotic scalings predicted by the above expression.

The bulk modulus follows from differentiation with respect to the
tension, $B^{-1}=\partial_\tau \epsilon$. The temperature induced
change is
\begin{equation}
  \label{eq:12}
  B^{-1}-B_a^{-1}\sim -\partial_\tau \langle\delta R^2\rangle\sim \left\{T \delta z^{-3} \qquad \tau\delta z^{-2}\ll1\atop T
       \tau^{-3/2} \qquad \tau\delta z^{-2}\gg1 \right. \;.
\end{equation}
The predicted cusp in the bulk modulus is reproduced in simulations,
as shown in Fig.~\ref{Fig:BulkScaling}.

\subsection{Shear modulus and fixed volume ensemble}
\label{sec:fixed-volume}
The zero-temperature shear modulus of jammed systems is soft close to isostaticity, as has been
extensively studied in the literature~\cite{wyart2005effects,tighe2011relaxations}. The characteristic shear modulus
\begin{equation}
  \label{eq:23}
  G\sim\left\{\delta z \qquad
    \tau\delta z^{-2}\ll1\atop  
       \tau^{1/2} \qquad \tau\delta z^{-2}\gg1  \right.
\end{equation}
is a direct consequence of the above spectrum and density of soft modes. 
% One way of rationalizing this behavior is to say that to shear the network by  an angle $\alpha$, all soft modes are excited by the same typical energy $e(\alpha)$ such that the mean square displacement scales as $\langle \delta R^2\rangle\sim \alpha \lmax$. A random walk of $\lmax$ displacements will then generate a displacement $\alpha \lmax$ on the scale of the correlation length, as required by the external shear. \oskar{The last paragraph may not be needed, but this is the way I rationalize the shear modulus.}

If the volume is kept fixed, temperature will generate a nonzero tension, which may stiffen the
shear modulus. The tension generated can be estimated by setting
$\epsilon=0$ (or more generally to a given pre-described value) in
(\ref{eq:11-0}),
\begin{equation}
  \label{eq:19}
  \tau_{\epsilon=0}\sim \left\{T \delta z^{-1} \qquad T\ll \delta z^{3} \atop 
  T^{2/3}  \qquad T\gg \delta z^{3} \right. \;.
\end{equation}
Inserting this scaling behavior for the tension into (\ref{eq:23}), we get
\begin{equation}
  \label{eq:24}
  G\sim\left\{\delta z \qquad
    T\ll \delta z^{3}\atop  
       T^{1/3} \qquad T\gg \delta z^{3}  \right.  \;.
\end{equation}
We thus obtain the scaling $G\sim T^{1/3}$ in the critical regime, as was shown in Fig.~\ref{fig:square}.

\section{Discussion}
\label{sec:Discussion}
We have shown that the elastic properties of near-isostatic spring networks derived from jammed packings of elastic spheres exhibit critical behavior upon heating. At fixed finite tension, the negative thermal expansion coefficient (NTE) diverges as $\tau^{-1/2}$ at isostaticity ($\delta z=0$). At vanishing tension, networks shrink with a NTE diverging as $\delta z^{-1}$. Concomitantly, the bulk modulus exhibits a cusp at criticality.

\begin{figure}[!tb]
\begin{center}
\includegraphics[width=0.5\textwidth]{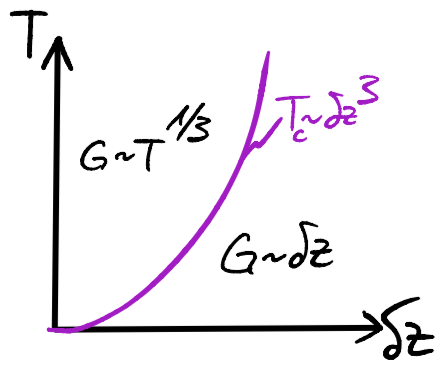}
\caption{\label{fig:Phasediagram}
\textbf{Effect of thermal fluctuations on the shear modulus at fixed volume.}
In this schematic phase diagram of the shear modulus, we indicate the phase boundary $|\delta z|\sim T^{-1/3}$ separating the zero-temperature behavior $G\sim\delta z$ from the critical regime $G\sim T^{1/3}$. The temperature controlled regime widens rapidly as temperatures are increased. \oskar{I guess we cannot make any statements about negative $\delta z$ unless we measure them.}}
\end{center}
\end{figure}

At fixed volume, the tendency to shrink leads to a rapid build-up of tension. As a consequence, the shear modulus at criticality stiffens with temperature as $G\propto T^{1/3}$. Our arguments predict that the region of stability is wider in heated networks, as illustrated in the phase diagram of Fig.~\ref{fig:Phasediagram}: As the temperature increases tension builds up to stabilize the network down to $-\delta z\sim T^{1/3}$. \oskar{(Can we even prepare networks with negative $\delta z$?)}

It is useful to compare our results  with those of Dennison et al.~\cite{dennison2013fluctuation}, in which spring networks at fixed volume close to isostaticity were also studied. The authors found that the shear modulus at criticality increases with temperature as $G\sim T^{1/2}$. This intriguing result is in contrast with our setup, which as mentioned exhibits $G\sim T^{1/3}$.

\begin{figure}[!tb]
\begin{center}
\includegraphics[width=0.5\textwidth]{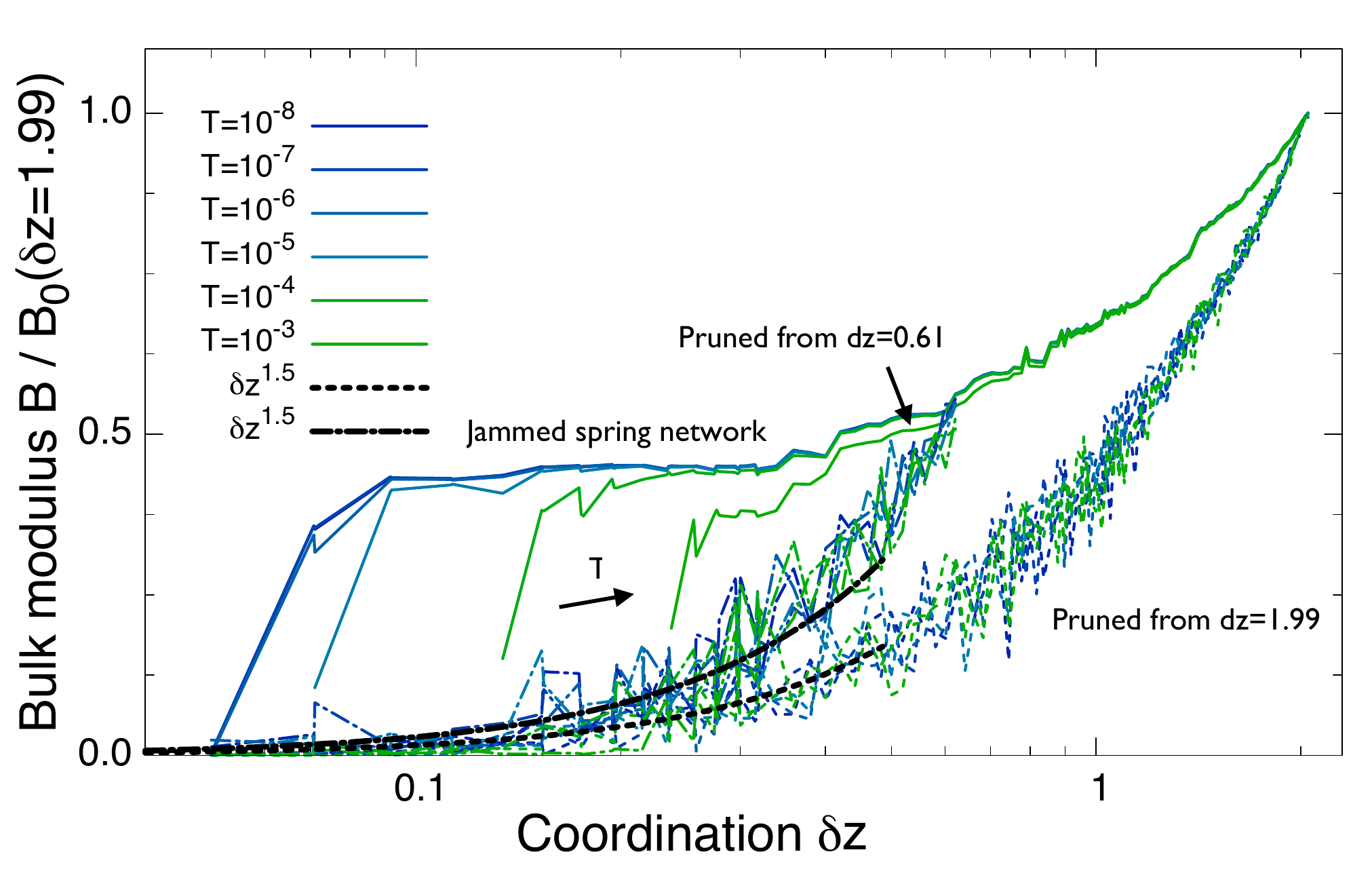}
\caption{\label{fig:Structure}
\textbf{Effect of structure on bulk modulus.
Bulk modulus for networks from packings(solid line) and their pruned counterparts where pruning started from $\delta z = 1.99$(dashed line) and $\delta z = 0.61$(solid and dotted line).}
The bulk modulus softens with $T$ for the networks from packings while it is temperature-independent for the pruned networks.
The upper bound of the bulk modulus is given by the networks from jammed packings when considered as a variational problem.
We see that the pruned networks do not couple to the tension like the networks from jammed packings do.
%Data as a function of the distance from the isostatic point of the jammed packing spring networks.
The bulk modulus is compared to its value at zero-temperature far away from the isostatic point $\delta z = 1.99$.
Note that the bulk modulus of pruned networks is well-described by a scaling $\propto
\delta z^{1.5}$. The same scaling behavior was observed for shear
moduli in the pruned networks of Ref.~\cite{wyart2008elasticity} close to isostaticity.
In contrast, the networks for jammed packings maintain their bulk modulus approaching the isostatic point unless they get softened by temperature.
In the limit of zero temperature the networks obtained from jammed packings have a large bulk modulus while the one of pruned networks vanishes, also see \cite{ellenbroek2009non}.
The wiggly data for pruned networks stems from every $\delta z$ data point being pruned independently. 
%Hence, on a qualitative scale we can say that the random component which induces variability in the pruning algorithm plays no important role.
System size is $N=100$.}
\end{center}
\end{figure}

This discrepancy indicates that the elastic properties of spring networks not only depend on the co-ordination number but also on the way that networks are prepared, as has been previously observed at zero temperature~\cite{ellenbroek2009non}. Dennison et al. started with highly co-ordinated networks and gradually removed springs until a given coordination number was achieved. These randomly diluted networks, which we call "pruned", are in contrast to our networks, which were generated at given co-ordination numbers directly from simulated particle packings. The comparison of the bulk moduli in both  network types in Fig.~\ref{fig:Structure} shows that pruned networks exhibit a soft bulk modulus for all temperatures in contrast to jammed networks, which have a bulk modulus of order the spring constant at vanishing temperature.

\oskar{(We could also try to discuss Mao et al, which have a $G\sim T^{2/3}$ scaling in the regime where thermal fluctuations stabilize their lattices. They also have a negative thermal expansion coefficient.)}

Finally, we hypothesize that heating in close-to-isostatic networks is somewhat analogous to shearing the network by a shear angle $\alpha\sim T^{1/2}$: In other words, we are assuming that all available floppy modes are excited by roughly the same amount of energy, proportional to $\alpha^2$. This shear-equipartition assumption for soft modes is consistent with existing scaling analyses, in particular Ref.~\cite{wyart2008elasticity}. Moreover, the so far unexplained temperature scaling $G\sim T^{1/2}$ for pruned networks then becomes  simply  a consequence of the scaling $G\sim \alpha$ for isostatic pruned networks at zero tension~\cite{wyart2008elasticity}. Extended to the shearing of \emph{jammed}, rather than pruned, isostatic networks, we moreover recover  the scaling of the negative shear dilatancy observed for fixed pressure in Ref.~\cite{tighe2014shear}. At finite volume, on the other hand, we predict that shearing jammed networks will lead to a sharp rise $G\sim \alpha^{2/3}$ of  shear modulus with shear angle, contrasting with the linear rise in pruned networks~\cite{wyart2008elasticity}. 

Ultimately, the response of our thermalized networks to tension (pressures)  results from the phonon  spectrum being dominated by many soft modes with frequencies that sensitively depend on tension. Similar mode spectra arise in (semiflexible) biopolymer networks, which therefore also exhibit  negative thermal expansion coefficients and shear dilatancy~\cite{janmey2007negative,conti2009cross,heussinger2007nonaffine}. 

% Also somewhere we should cite work on isostatic fiber networks \cite{broedersz2011criticality,feng2016nonlinear}, perhaps at the same place.

Our study focused on spring networks without the possibility of change in connectivity, in particular by bond breaking. A jammed random close pack of elastic spheres simply has no space to contract. Hence, heating of such packings without free space must immediately lead to bond breaking to ensure a mode spectrum consistent with the density constraint. Contraction as observed in our simulation may occur if a packing is over-coordinated (i.e. due to compression) or exhibits enough free volume and some degree of attractive interactions. We  suspect that the mean square displacement is the  quantity most robustly observed in packings, as it does not rely on the precise coupling of external forces and local tension.

% \begin{figure}[!tb]
% \begin{center}
% \includegraphics[width=0.5\textwidth]{figures/MSD_Pack.pdf}
% \caption{\label{fig:PackingMSD}
% \textbf{Scaling  with $\delta z$ of the Mean Square Displacement (MSD) in packings of elastic spheres.}
% On short enough times, where big rearrangements of the connectivity matrix are improbably, our simulations recover the scaling $MSD \sim T\delta z^{-1}$ observed for the corresponding spring networks.}
% \end{center}
% \end{figure}

\section*{Acknowledgments}
We would like to thank Matthieu Wyart for
helpful discussions. Research reported in this publication was supported by the National Institute Of General Medical Sciences of the National Institutes of Health under Award Number R01GM115851, by a Simons Investigator award from the Simons
Foundation (O.H.) and by the German Research Foundation (DFG) in the
framework of the SFB 937/A15 (OH, CB), SFB 937/A16 (CH, CB) and of the Emmy Noether  Program He 6322/1-1 (CH, CB). The content is solely the responsibility
of the authors and does not necessarily represent the official views
of the National Institutes of Health.

\bibliography{jamming}

\end{document}